%% file: main.tex
\documentclass[runningheads]{llncs}

\usepackage[T1]{fontenc}
\usepackage{newtxtext}
\usepackage[varvw]{newtxmath}
\usepackage{graphicx}
\usepackage{booktabs}
\usepackage{multirow}
\usepackage{makecell}
\usepackage{siunitx}
\usepackage{csquotes}
\usepackage{comment}
\usepackage{glossaries}
\usepackage[hidelinks]{hyperref}
\usepackage{cleveref}

\input{glossary}

\begin{document}

\title{An Empirical Evaluation of Validation Practices for Machine Learning–Based Network Intrusion Detection}
\titlerunning{Validation Practices for Machine Learning–Based Network Intrusion Detection}
\author{João R. Campos\\\email{jrcampos@dei.uc.pt}}
\authorrunning{J. R. Campos}
\institute{University of Coimbra, \\CISUC/LASI -- Centre for Informatics and Systems of the University of Coimbra, \\Department of Informatics Engineering, Coimbra, Portugal
}

\maketitle

\begin{abstract}
\gls{ml} techniques have attracted significant research interest in this domain, with numerous studies reporting near-perfect results on widely used intrusion datasets. In this practical experience report, we conduct a systematic evaluation as a reality check of current ML-based intrusion detection approaches reported in the literature, focusing on widely used public datasets under more careful evaluation conditions. We first reproduce representative prior studies using commonly adopted validation practices, highlighting how easily reported results can become biased. The same models are then reevaluated under stricter, leakage-resistant conditions, revealing that performance is frequently inflated by improper partitioning and weak cross-dataset generalization. Model behavior is analyzed across binary, multiclass, and expert detection tasks, as well as under cross-dataset validation, to identify limitations under more realistic assumptions. Overall, our results indicate that the effective detection capability of current ML-based IDS on public flow datasets appears to be substantially more limited than suggested by optimistic benchmark results, highlighting the need for caution when interpreting reported performance and for greater rigor in empirical evaluation.

\end{abstract}

\glsresetall 

\section{Introduction}
Modern organizations depend on increasingly complex software systems and web services, especially with hybrid and remote work. Greater complexity in architecture and codebases expands the attack surface, making systems harder to secure and more prone to defects. These vulnerabilities raise the risk of outages and data breaches, which can cause severe financial and reputational damage. Meanwhile, attackers are becoming more sophisticated, employing methods from DoS attacks to ransomware and data exfiltration \cite{enisa2023}. In 2022, the average data breach cost \$4.35 million, with global damages projected to reach \$10.5 trillion by 2025 \cite{Gadient2024}.

\glspl{ids} monitor systems or networks for abnormal or malicious activity, raising alerts to enable timely response. They can be deployed as network appliances or host-based agents, with some including automated mitigation (\glspl{ips}/\glspl{idps}). Signature-based systems like Snort and Suricata remain common, but they require prior knowledge of attacks and cannot detect zero-day threats. To address these challenges, recent research has increasingly integrated \gls{ml} techniques into intrusion detection. By learning behavioral patterns and distinguishing between normal and anomalous activity, \gls{ml} may help detect previously unseen threats. 

Despite its potential, \gls{ml}-based \glspl{ids} face several challenges. Algorithmic performance varies widely, methodological flaws can produce misleading results, and the domain requires expertise spanning both security and \gls{ml}. When this expertise is lacking, features may be misused, leading to overfitting and poor generalization. Most recent work has focused on \glspl{nids}, supported by widely available public datasets, with many studies reporting near-perfect detection performance on these benchmarks \cite{Elmrabit2020,more2024enhanced,putro2025evaluating,sharma2025improving,Kurniabudi2020,Liu2022}. In contrast, real-world intrusions remain frequent and are often detected only after weeks or months, raising doubts about the practical relevance of such optimistic results. The fact is that many published studies lack transparency or reproducibility, omitting critical details such as preprocessing steps, model architectures, or data partitioning strategies. This is particularly problematic for \gls{ml}-based IDS, where improper validation, especially naive partitioning and inadequate features, can easily introduce information leakage, leading to inflated performance that does not reflect realistic detection capability. As these benchmarks increasingly guide research direction and practitioner expectations, inflated or misleading evaluations risk reinforcing false confidence in ML-based intrusion detection and spreading poor evaluation practices.

This paper presents a critical and experience-driven evaluation of ML-based network intrusion detection using widely adopted public datasets. Rather than introducing new models or evaluation frameworks, our objective is to assess what level of performance can realistically be expected from state-of-the-art ML-based approaches with these datasets when common evaluation pitfalls, such as information leakage and inappropriate data partitioning, are avoided. By replicating representative prior setups and progressively applying more conservative evaluation conditions, we expose a substantial gap between the optimistic benchmark results reported in previous works and the detection capabilities observed under more realistic assumptions.

To support this analysis, we examine IDS performance across binary, multiclass, and expert classification settings, highlighting the strengths and limitations of each formulation. While binary classifiers often retain acceptable aggregate performance, multiclass models frequently fail to detect minority attack types, and expert models, although more effective for specific threats, introduce additional complexity and deployment trade-offs. We further evaluate cross-dataset generalization under shared feature and label constraints, showing that models trained on one benchmark dataset struggle to transfer even to closely related datasets, raising doubts about their robustness and reuse.

Beyond reporting performance figures, this work adopts a practice-oriented perspective on evaluating ML-based network intrusion detection using public datasets. We show that naive random splits can substantially inflate reported results, that commonly used aggregate metrics often obscure severe detection failures, and that benchmark-driven evaluations tend to overstate real-world effectiveness. By reassessing representative models under leakage-resistant validation conditions, we illustrate how performance degrades once these sources of bias are removed. These observations should be interpreted as guidance for setting realistic expectations and for conducting more rigorous empirical evaluations, rather than as claims of deployable IDS performance.


\section{Background and Related Work}
\label{chap:background}

\subsection{\glsentryfull{ids}}
\glspl{ids} monitor networks or hosts for malicious activity, alerting administrators or actively blocking threats (in which case they are called \glspl{idps})~\cite{Scarfone2007}. They may be hardware- or software-based and operate in either passive or active modes.

\glspl{ids} are commonly categorized by both their detection methodology and deployment strategy. \glspl{sids} rely on predefined rules to detect known threats~\cite{Khraisat2019}; they are accurate but require frequent updates and struggle with novel or distributed attacks, although tools like Snort and Suricata mitigate these limitations through traffic tracking and rule thresholds. \glspl{aids} instead learn normal behavior and detect deviations using statistics, heuristics, or \gls{ml}\cite{Khraisat2019}; while adaptable, they risk incorporating malicious activity into their baseline and often generate false positives in complex environments\cite{Scarfone2007}. From a deployment perspective, \glspl{nids} monitor traffic at gateways or routers~\cite{Scarfone2007}, proving effective at catching early-stage attacks but struggling in high-throughput networks~\cite{Bulajoul2015}, whereas \glspl{hids} run on individual systems, analyzing logs, processes, filesystem activity, and local traffic~\cite{Khraisat2019}; they are effective for insider threats and encrypted traffic but introduce resource overhead and may delay detection if not continuous~\cite{Scarfone2007}. A typical detection pipeline involves measurement, classification, and response.

Numerous datasets for assisting in the development of IDSs have been introduced, differing in realism, feature sets, and quality. Public datasets for IDS are dominated by \glspl{nids}, with \glspl{hids} datasets such as ADFA-LD~\cite{Creech2014} and ADFA-WD~\cite{Haider2016} now outdated and inaccessible.\footnote{\url{https://research.unsw.edu.au/projects/adfa-ids-datasets}}\footnote{\url{https://azsecure-data.org/other-data.html}} Most studies therefore rely on network flow datasets, where aggregated traffic features (packet sizes, flags, timing) are exported via NetFlow/IPFIX for scalable analysis~\cite{Hofstede2014}. Early benchmarks include KDD99~\cite{Tavallaee2009} and NSL-KDD~\cite{Tavallaee2009}, which remain widely cited despite synthetic traffic, redundancy, and imbalance. Kyoto 2006+\cite{Song2011a} and Twente\cite{Sperotto2009} captured honeypot traffic but suffer from labeling uncertainty and limited availability. ISCX-IDS2012~\cite{Shiravi2012} simulated structured traffic but lacks multiclass labeling. More recent datasets such as UNSW-NB15~\cite{Moustafa2015}, UGR’16~\cite{Macia-Fernandez2018}, CIC-IDS2017/2018~\cite{cicids2017,Sharafaldin2018}, and LITNET-2020~\cite{Damasevicius2020} offer richer features and larger scales, yet all present issues ranging from proprietary tools and incomplete documentation to inconsistent labeling and limited accessibility. Heterogeneous feature sets further complicate cross-dataset evaluation, as systems trained on one dataset rarely transfer without significant preprocessing. To mitigate this, Sarhan et al.~\cite{Sarhan2022} proposed NetFlow-compatible features for more consistent comparisons. Other resources, such as CAIDA DDoS 2007 and MAWILab, provide valuable packet-level data but require extensive preprocessing.

\subsection{Machine Learning}
\gls{ml} is central to modern \glspl{ids}, with supervised learning being a widely used paradigm (e.g., ~\cite{Yin2017}). Classifiers are trained on labeled data to distinguish normal from malicious activity, using common algorithms such as \glspl{svm}, or \glspl{nn}. Ensemble methods such as \gls{rf} can further improve accuracy and robustness by combining multiple learners.

A key limitation of supervised learning is its dependence on labeled datasets, which are costly, error-prone, and impractical at scale, especially for emerging threats. Unsupervised learning addresses this by modeling normal behavior and flagging deviations as anomalies, without requiring labeled attacks. Popular techniques include clustering (e.g., \textit{k}-Means, DBSCAN) and representation learning (e.g., \gls{pca}, AutoEncoders). While more flexible in dynamic environments, unsupervised methods are prone to false positives from rare but benign behavior. 

High dimensionality further complicates intrusion detection, increasing model complexity, training time, and overfitting risk. Feature selection or extraction and instance selection mitigate these issues by retaining the most informative features or representative samples. Class imbalance can bias models toward majority classes. Undersampling and oversampling can reduce this bias.

Model evaluation depends strongly on data partitioning. While strategies such as $k$-fold cross-validation, stratification, or bootstrap resampling are commonly used to improve reliability~\cite{arlot2010survey}, temporal dependence between observations complicates their application. In sequential domains, out-of-sample evaluation is often adopted to prevent using “future” data to predict the “past”\cite{tashman2000out}, but this approach limits data usage and typically allows only a single evaluation. Rolling-origin variants partially address this limitation, yet there remains no consensus on appropriate validation strategies for temporally correlated data, with some studies arguing that traditional cross-validation can still yield reliable estimates in practice\cite{bergmeir2018note}. These latter approaches are frequently applied in IDS research. In IDS the problem is exacerbated by the structure of network traffic datasets, where attacks are often executed as campaigns that generate highly correlated traffic and, subsequently, flows. As a result, random or shuffled splits are particularly prone to information leakage, leading to overly optimistic performance estimates. To mitigate similar issues in dependability-related prediction tasks, \gls{eloocv} was proposed~\cite{Campos2023}, which treats each attack or experiment as an atomic unit and testing only on entirely unseen events, preventing leakage between correlated samples. Although originally introduced for operating system failure prediction, it naturally generalizes to intrusion detection and has already been applied in IDS contexts~\cite{castro2025developing}.

Metric selection is equally crucial: accuracy can mask poor detection of minority classes, so measures based on the \gls{cm} such as precision, recall, F1, and areas under \gls{roc} or \gls{pr} curves are preferred. Metrics that weight all classes or emphasize class-specific recall are particularly important in imbalanced domains such as intrusion detection~\cite{Sokolova2009}.

\subsection{Related Work}

ML-based \gls{nids} has been studied for several years, with most empirical work grounded in public benchmark datasets. A recurring characteristic across this literature is the reporting of consistently high performance. 

For UNSW-NB15, More et al.\ \cite{more2024enhanced} report over 98\% accuracy and an F1-score of 0.98 in a binary classification setting using standard classifiers, including logistic regression, SVM, decision trees, and random forests. Putra \cite{putra2024evaluating} presents a comparative evaluation in which a neural network achieves 92\% accuracy, while SVM and k-NN reach 90\% and 88\%, respectively. Moualla et al.\ \cite{Moualla2021} further improve performance on UNSW-NB15 by applying SMOTE-based oversampling and feature selection, reporting strong results across multiple classes with near-perfect performance, although with noticeable variation in recall for certain attack types, particularly DoS. Azeroual et al.\ \cite{azeroual2022analysis} combine random forests, XGBoost, and decision trees with PCA-based feature reduction and report strong results, including a reported test accuracy of 100\% for a decision tree classifier under a random split. In a broader comparative study, Putro \cite{putro2025evaluating} reports that Random Forest achieves a 95\% test accuracy on UNSW-NB15, while also noting substantially higher training scores for several models.

A comparable tendency toward strong reported performance is observed for the CIC family of datasets. Elmrabit et al.\ \cite{Elmrabit2020} evaluate multiple classical machine learning algorithms on CIC-IDS2017 and report near-perfect performance for several classifiers under standard holdout and cross-validation splits, including precision and recall values exceeding 99\%. Feature-selection and benchmarking studies similarly report very high headline results. Kurniabudi et al.\ \cite{Kurniabudi2020} apply Information Gain-based feature selection on CIC-IDS2017 and report over 99\% accuracy for Random Forest and  J48, using reduced feature subsets. Sharma et al.\ \cite{sharma2025improving} report that a hybrid CapsNet–BiLSTM model achieves 99\% accuracy, with precision, recall, and F1-scores close to 0.99. Maseer et al.\ \cite{maseer2021benchmarking} also report accuracies above 99\% for several classifiers, although these results correspond to a subset of the classes. Yang et al.\ \cite{yang2025high} report that a multi-scale CNN achieves accuracy exceeding 99\% on CIC-IDS2017 and surpassing 99\% on UNSW-NB15.

For the CSE-CIC-IDS2018 dataset, similarly strong performance is commonly reported. Farhan and Jasim \cite{farhan2022performance} evaluate a deep learning–based intrusion detection model using LSTM architectures and report detection accuracies close to 99\% across multiple attack categories. Elhanashi et al.\ \cite{elhanashi2022machine} present a comparative evaluation of classical machine learning techniques on the same dataset, including Random Forest, decision trees, and support vector machines, reporting high overall accuracy and detection rates for several classifiers. 

A common characteristic across many of these studies is the use of relatively simple data partitioning strategies, such as fixed 70/30 or 80/20 train–test splits \cite{Elmrabit2020,Kurniabudi2020,Moualla2021}, or shuffled holdout validation \cite{Farhat2023}, often without explicit consideration of temporal dependencies or attack-level structure. In the presence of attack campaigns that generate highly correlated flows—such as \texttt{Slowloris} or SSH brute-force attacks—random or shuffled splits may distribute related samples across training and test sets. This practice can introduce information leakage and substantially inflate reported performance, leading to overly optimistic estimates of detection capability.

Apruzzese et al.~\cite{Apruzzese2022XeNIDS} introduced XeNIDS to evaluate flow-based intrusion detectors across network contexts, including ensembles of attack-specific detectors. Their within-context baseline used random 80/20 splits at the sample level. Our within-dataset analysis compares conventional flow-level splits with whole-experiment holdouts to assess how keeping correlated flows together affects measured performance. Verkerken et al.~\cite{Verkerken2022Generalization} found that four unsupervised methods performed well within individual datasets but lost performance when transferred between CIC-IDS2017 and CSE-CIC-IDS2018. Our transfer experiments provide complementary evidence for supervised classifiers on the same dataset pair.

\section{Methodology}
\label{chap:methodology}
The development and evaluation of effective predictors are strongly influenced by the characteristics of the problem domain. In the context of ML-based network intrusion detection, these characteristics determine how data should be prepared, which learning techniques are appropriate, and how models should be evaluated. Such choices directly affect not only predictive performance, but also the validity and interpretability of the resulting conclusions.


Besides improper partitioning strategies, insufficient consideration of dataset characteristics can also lead to inappropriate evaluation choices. The use of aggregate metrics alone can obscure critical failure modes, particularly in highly imbalanced settings. In intrusion detection, minimizing the false negative rate is essential, as undetected attacks may have severe consequences, while excessively high false positive rates can overwhelm operators and render a system impractical. Balancing these trade-offs requires evaluation protocols that reflect realistic operational conditions.

Motivated by the consistently high performance reported in prior work, we aim to reassess these results under progressively stricter evaluation conditions. Our methodology begins by reproducing representative evaluation setups from the literature and then systematically refines data partitioning and validation strategies to reduce information leakage and better reflect realistic deployment scenarios.

\subsection{Network Intrusion Datasets}
\label{sec:methodology_datasets}
For this work we selected four widely used flow-based intrusion detection datasets: UNSW-NB15 (2015, 2.5M samples, 48 features, 10 classes), CIC-IDS2017 (2017, 2.8M samples, 78 features, 15 classes), CIC-IDS2018 (2018, 16M samples, 79 features, 15 classes), and the corrected CNS-IDS2017 dataset (2022, 2.8M samples, 91 features, 15 classes). We selected these datasets for their widespread use, diverse traffic and features, and suitability for evaluating cross-dataset transfer.

UNSW-NB15 remains one of the most commonly used benchmarks, despite its relatively small size, pronounced class imbalance (approximately 7:1 benign-to-attack ratio), and partial dependence on proprietary traffic generation and parsing tools. The CIC-IDS datasets provide more recent traffic traces and a wider range of attack scenarios, but are known to suffer from issues in flow extraction and labeling introduced by CICFlowMeter, including segmentation errors and mislabeled flows. These limitations were addressed by Liu et al.~\cite{Liu2022}, who introduced the CNS-IDS datasets by reprocessing the raw traffic with improved flow generation and corrected labels, resulting in more reliable benchmarks. We used CNS-IDS2017 and excluded CNS-IDS2018 (approximately 63M samples, 33~GB) due to compute and storage constraints.

We also evaluated standardized NetFlow-based feature variants proposed by Sarhan et al.~\cite{Sarhan2022}, which aim to harmonize feature sets across datasets to facilitate transfer learning. However, these representations aggregate all flows into a single CSV file without preserving temporal boundaries, making them unsuitable for our evaluation methodology. Since our approach relies on experiment-based partitioning to prevent information leakage, preserving temporal structure is a strict requirement.

\subsection{Evaluation and Algorithms}

Data partitioning critically affects the evaluation of ML-based intrusion detection systems, shaping both performance estimates and susceptibility to overfitting. In flow-based \glspl{nids} datasets, flows are not independent: a single attack campaign can generate many correlated flows. Random shuffling and splitting therefore risks distributing related samples across training and test sets, introducing information leakage. This can inflate estimated performance for attacks such as \texttt{Slowloris} and SSH brute-force.

To replicate prior work and establish a baseline, we first evaluated models using traditional stratified 10-fold cross-validation, following common practice in the literature. We then adopted a stricter and more representative evaluation strategy based on \gls{eloocv}~\cite{Campos2023}. Under this approach, each attack campaign is treated as an atomic unit: models are trained on all but one campaign and tested on the held-out campaign, preventing leakage even when attacks generate large numbers of correlated flows. Since many public datasets lack explicit campaign identifiers, some attack groups needed to be approximated using daily boundaries, ensuring that flows from the same attack never appear in both training and test sets.

This choice introduces trade-offs. Attack classes executed on a single day cannot be evaluated as standalone supervised classes without violating the closed-set assumption (you need at least two experiments of each label/attack). When conceptually appropriate, we addressed this limitation through label coalescing, grouping semantically similar attack variants (e.g., DoS tools) into a single class, following common practice in prior work (e.g., \cite{Kurniabudi2020}). When coalescing was not feasible, the affected labels were excluded from multiclass evaluation. In some CIC datasets, this results in held-out days containing tool-specific implementations not observed during training, particularly for the DoS and Patator families in CIC-IDS2017 and the DoS/DDoS classes in CIC-IDS2018. We interpret this behavior as a conservative test of generalization at the level of attack categories rather than as a violation of supervised learning assumptions.

All experiments were conducted using the Propheticus framework~\cite{Campos2019}, which provides an end-to-end, reproducible ML pipeline from preprocessing to evaluation. The experiment configurations and preprocessing scripts will be released in a reproducibility package \footnote{\url{https://figshare.com/s/be012bb1646ccbdf766e}}. To enable comparison with existing literature, we included baseline models such as logistic regression. For the main evaluation, we focused on \gls{rf} and \gls{xgboost}, given their strong performance in prior IDS studies and their effectiveness on tabular data. Hyperparameters were tuned via grid search.

Evaluation relied on precision, recall, F1-score, balanced accuracy, and area under the precision–recall curve (AUPRC), which better reflect performance in highly imbalanced settings than accuracy or ROC AUC. To mitigate class imbalance during training, random undersampling was applied to the majority class, while test distributions were left unchanged.

We evaluated three complementary tasks: (i) binary and multiclass classification under \gls{eloocv}, contrasted with naive 70/30 splits~\cite{Elmrabit2020} to quantify the impact of information leakage; (ii) cross-dataset generalization, training on one dataset and testing on another (e.g., CIC-IDS2017 to CIC-IDS2018) using only shared features and labels; and (iii) expert models, implemented as binary classifiers specialized for individual attack types, to assess per-attack detection capability under realistic conditions.

\subsection{Data Preparation}

Dataset-specific preprocessing focused on removing IP-based features, eliminating missing or infinite values, and encoding categorical variables in a binary format. To reduce bias, we removed features that could readily leak labels or encode tool-specific artifacts.

\textit{UNSW-NB15:}  
The full dataset (2.5M samples) was used after correcting label inconsistencies, missing values, and formatting issues such as trailing whitespace. Redundant attributes (e.g., \texttt{label\_bin}, \texttt{ct\_state\_ttl}) and leakage-prone features (e.g., \texttt{is\_ftp\_login}, \texttt{ct\_ftp\_cmd}) were removed. Placeholder values were standardized, and categorical features were binary-encoded.

\textit{CIC-IDS2017:}  
Preprocessing addressed duplicate \texttt{Fwd Header Length} fields, malformed Unicode entries, and inconsistent labels. To preserve sufficient samples for evaluation under experiment-based/day-based \gls{eloocv} and to maintain closed-set supervised learning, attack classes executed on a single day were coalesced following common practice in the literature (e.g., \cite{Kurniabudi2020}). In CIC-IDS2017, all DoS attacks occur on a single day and are implemented using different tools (HULK, GoldenEye, Slowloris, SlowHTTPTest, and Heartleech). These were merged into a single \textit{DoS} class, where each attack represents a different experiment. Similarly, SSH and FTP Patator attacks were merged into a single \textit{Patator} class, and \textit{PortScan} was merged into \textit{Infiltration}. Although these tools differ in implementation, CICFlowMeter does not encode tool- or protocol-specific semantics, and their variants collapse into similar flow-level statistics. Our focus is on detecting attack types rather than individual tools.

\textit{CIC-IDS2018:}  
CSV files were cleaned to remove duplicates and metadata errors. The \textit{Benign} class was randomly undersampled from approximately 13.5M to 2M samples to reduce class imbalance. Attack coalescing followed the same strategy as CIC-IDS2017 to ensure consistency across datasets, and cross-dataset experiments were restricted to shared feature sets. As in CIC-IDS2017, all DoS attacks occur on a single day and rely on similar tools. For DDoS attacks, LOIC and HOIC generate high-rate HTTP floods that result in comparable flow-level behavior. SSH and FTP Patator attacks could not be separated due to flow capture limitations and were therefore treated as a single class, preventing their independent evaluation under \gls{eloocv}.

\textit{CNS-IDS2017:}  
The dataset was reprocessed by Liu et al.~\cite{Liu2022} to correct labeling and flow segmentation errors. Metadata fields were removed, and only successful attack instances were retained for analysis.

Attack classes for which only a single experiment was available could not be evaluated as standalone classes under \gls{eloocv} and were excluded from multiclass evaluation when necessary. Where possible, we retained these classes for cross-dataset validation.

The final sample distribution for all datasets is summarized in Table~\ref{tab:distribution}.

\begin{table}[!t]
	\small
	\begin{center}
		\caption{Final data distribution of chosen datasets}\label{tab:distribution}
		\begin{tabular}[c]{@{}llrr@{}}
			Dataset     & Label            & \#Samples & \#Experiments \\ \midrule
			UNSW-NB15   & Benign           & 2218764   & 4             \\
			& Analysis         & 2677      & 4             \\
			& Backdoors        & 2329      & 4             \\
			& DoS              & 16353     & 4             \\
			& Exploits         & 44525     & 4             \\
			& Fuzzers          & 24246     & 4             \\
			& Generic          & 215481    & 4             \\
			& Reconnaissance   & 13987     & 4             \\
			& Shellcode        & 1511      & 4             \\
			& Worms            & 174       & 4             \\ \cmidrule{1-4}
			CIC-IDS2017 & Benign           & 2271320   & 8             \\
			& Bot              & 1956      & 1             \\
			& DDoS             & 128025    & 1             \\
			& DoS              & 251723    & 5             \\
			& Patator          & 13832     & 2             \\
			& Infiltration     & 158840    & 2             \\
			& Web-Bruteforce   & 1507      & 1             \\
			& Web-SQLi         & 21        & 1             \\
			& Web-XSS          & 652       & 1             \\ \cmidrule{1-4}
			CIC-IDS2018 & Benign           & 1986024   & 10             \\
			& Bot              & 286191    & 1             \\
			& DDoS             & 1263933   & 3             \\
			& DoS              & 654300    & 4             \\
			& Patator          & 380943    & 1             \\
			& Infiltration     & 160639    & 2             \\
			& Web-Bruteforce   & 611       & 2             \\
			& Web-SQLi         & 87        & 2             \\
			& Web-XSS          & 230       & 2             \\ \cmidrule{1-4}
			CNS-IDS2017 & Benign           & 1594540   & 5             \\
			& Bot              & 736       & 1             \\
			& DDoS             & 95144     & 1             \\
			& DoS              & 171645    & 5             \\
			& Patator          & 6933      & 2             \\
			& Infiltration     & 230869    & 3             \\
			& Web-Bruteforce   & 73        & 1             \\
			& Web-SQLi         & 13        & 1             \\
			& Web-XSS          & 18        & 1             \\
		\end{tabular}
	\end{center}
\end{table}

\section{Results and Discussion}
\label{chap:results}

\sisetup{
	round-mode=places,
	round-precision=3,
	text-series-to-math=true,
	propagate-math-font=true
}
This section presents the experimental results under progressively stricter evaluation conditions, highlighting the limitations of commonly used practices and motivating the need for more careful validation. In \Cref{sec:replicating_previous_work}, we begin with standard evaluation setups to establish a baseline, replicating representative results from prior work and verifying reported performance. This analysis reveals methodological issues, most notably information leakage arising from improper data partitioning. To mitigate these effects, \Cref{sec:developing_representative_predictors} introduces a stricter \gls{eloocv}-based evaluation strategy, which yields lower but more realistic performance estimates. Finally, \Cref{sec:evaluating_generalization} examines cross-dataset generalization between CIC-IDS2017 and CIC-IDS2018, highlighting the limited transferability of models across changing deployment conditions even under similar attacks.

\subsection{Replicating Previous Work}\label{sec:replicating_previous_work}
To assess the impact of more conservative partitioning, we began by evaluating IDS models under standard data partitioning approaches commonly used in related work. This provided a baseline under widely used methodologies and helped identify the sources for inflated performance. We reproduced common setups, combining two partitioning strategies (70/30 split and 10-fold stratified cross-validation) with four algorithms, \gls{lr}, \gls{dt}, AdaBoost, and \gls{rf}.

Datasets were evaluated without balancing; features were normalized using \textit{z-score}; and low-variance features were removed, aligned with current practices also observed in related works. For both 70/30 and 10-fold CV shuffling and stratification was used. Since prior works rarely specify hyperparameters, we used defaults, modifying only \gls{lr} maximum iterations to ensure convergence. Our intent was not fine-tuning but to create standard predictors and establish a baseline, while allowing us to reproduce performance claims of related works under comparable conditions.

\subsubsection*{UNSW-NB15}
Under the 70/30 split, \gls{rf} achieved binary and multiclass F1-scores of 0.983 and 0.872, compared with 0.912 and 0.695 reported by Elmrabit et al.~\cite{Elmrabit2020}.
All four classifiers had higher binary than multiclass F1-scores.

\subsubsection*{CIC-IDS2017}
Results on CIC-IDS2017 showed an unexpected pattern: most classifiers improved when moving from binary to multiclass tasks. For the binary setting, \gls{dt} achieved 99.6\% precision and recall, which improved slightly in the multiclass case, similar to other works in the literature. A direct comparison between the results reported by Elmrabit et al.\ and our baseline replication is shown in Table~\ref{tab:prev_work_2017}, showing that conventional splits can reproduce near-perfect scores.

{\setlength{\emergencystretch}{2em}
The prevalence of such extreme scores is consistent with prior work (e.g., \cite{Elmrabit2020,Kurniabudi2020,maseer2021benchmarking,sharma2025improving}), which likewise reports very high performance, often approaching perfection. This underscores a central problem: if taken at face value, such results would suggest that intrusion detection is essentially solved, which is clearly contradicted by ongoing attacks and repeated warnings from agencies and annual threat reports~\cite{enisa2023}. These inconsistencies point not to genuine model capability, but to methodological flaws or issues in the data. To evaluate this, we shift to the more rigorous \gls{eloocv}~\cite{Campos2023} approach, which isolates attack campaigns across train-test splits and provides a more realistic measure of model generalization.\par}

\begin{table}[!t]
	\centering
\caption{Comparison between previously reported results and our baseline replication for CIC-IDS2017}
\begin{minipage}{\linewidth}
		\centering
		\small
		\textbf{(a) Results from Elmrabit et al.}\par\smallskip 			\setlength{\tabcolsep}{3pt} 

		\begin{tabular}{@{}llrrr@{}} 
			Binary & Algorithm          & Precision & Recall & F1-Score \\ \midrule
			Yes    & LogisticRegression & 0.737     & 0.634  & 0.682 \\
			& DecisionTree       & 0.995     & 0.996  & 0.996 \\
			& AdaBoost           & 0.898     & 0.910  & 0.904 \\
			& \gls{rf}      & 0.997     & 0.997  & 0.997 \\ \cmidrule{1-5}
			No     & LogisticRegression & 0.914     & 0.915  & 0.910 \\
			& DecisionTree       & 0.998     & 0.998  & 0.998 \\
			& AdaBoost           & 0.769     & 0.818  & 0.760 \\
			& \gls{rf}      & 0.999     & 0.999  & 0.999 \\
		\end{tabular}
	\end{minipage}\par\medskip
	\begin{minipage}{\linewidth}
		\centering
		\small
		\textbf{(b) Our results}\par\smallskip 
			\setlength{\tabcolsep}{3pt} 

		\begin{tabular}{@{}lllrrr@{}}
			Binary & Partion.    & Algorithm          & Precision & Recall & F1-Score \\ \midrule
			Yes    & 70/30           & LogisticRegression & 0.846     & 0.784  & 0.814    \\
			&                 & DecisionTree       & 0.997     & 0.997  & 0.997    \\
			&                 & AdaBoost           & 0.969     & 0.981  & 0.975    \\
			&                 & \gls{rf}      & 0.996     & 0.997  & 0.997    \\ \cmidrule{2-6}
			& \multirow{2}{*}{\makecell{Stratified \\ KFold}} & LogisticRegression & 0.921 & 0.962 & 0.941 \\
&                                                & DecisionTree       & 0.999 & 0.999 & 0.999 \\
			&                 & AdaBoost           & 0.991     & 0.987  & 0.989    \\
			&                 & \gls{rf}      & 0.999     & 0.999  & 0.999    \\ \cmidrule{1-6}
			No     & 70/30           & LogisticRegression & 0.956     & 0.959  & 0.956    \\
			&                 & DecisionTree       & 0.998     & 0.998  & 0.998    \\
			&                 & AdaBoost           & 0.757     & 0.857  & 0.804    \\
			&                 & \gls{rf}      & 0.998     & 0.998  & 0.998    \\
		\end{tabular}
	\end{minipage}

	\label{tab:prev_work_2017}
\end{table}

\subsection{Developing Representative Predictors}\label{sec:developing_representative_predictors}

Unlike the earlier experiments which were limited to UNSW-NB15 and CIC-IDS2017, we extended our evaluation to include the datasets introduced in \Cref{sec:methodology_datasets}, allowing for a broader and more rigorous assessment of model performance.

\subsubsection*{UNSW-NB15}
Under \gls{eloocv} with a maximum tree depth of 7, balanced accuracy fell from 99.3\% for binary classification to 38.2\% for multiclass classification with \gls{rf}, and from 96.9\% to 25.4\% with \gls{xgboost}.

\begin{figure}[!htbp]
	\centering
	\includegraphics[width=0.70\linewidth]{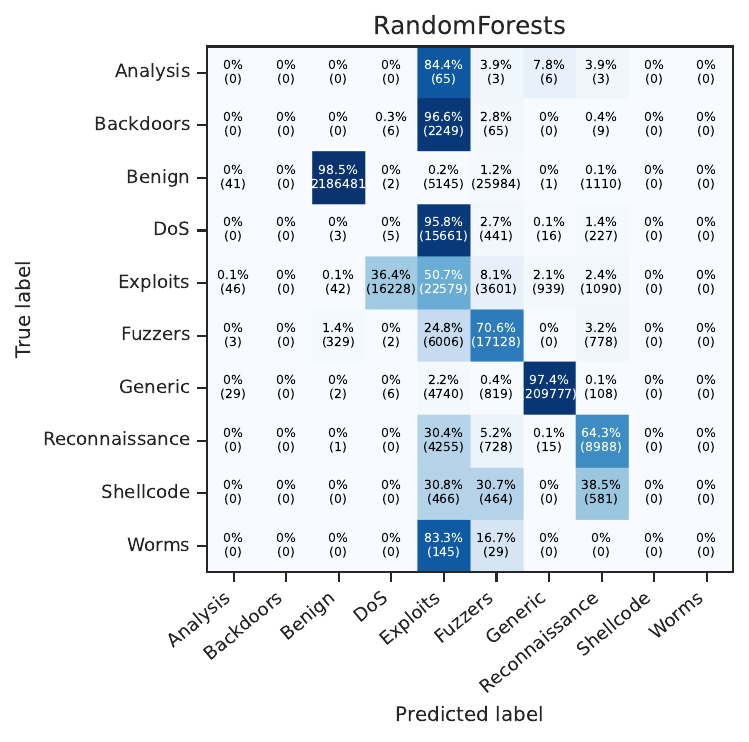}
	\caption{UNSW-NB15 Multiclass}
	\label{fig:unsw_cm_multi}
\end{figure}

Weighted multiclass recall remained higher, at 80.5\% for \gls{rf} and 68.0\% for \gls{xgboost}, because common classes carry more weight. As shown in \Cref{fig:unsw_cm_multi}, only \textit{Benign} and \textit{Generic} were consistently classified correctly. Many \textit{Analysis}, \textit{DoS}, and \textit{Shellcode} flows were misclassified as \textit{Exploits}.

These observations can largely be attributed to the dataset’s extreme imbalance. As noted in \Cref{tab:distribution}, several classes have fewer than 1\% of the total samples, making it difficult for general-purpose classifiers to learn meaningful patterns from them, even under balanced benign training. 

\subsubsection*{CIC-IDS2017}

\begin{figure}[!htbp]
	\centering
	\begin{minipage}[t]{0.49\linewidth}
		\centering
		\includegraphics[width=\linewidth]{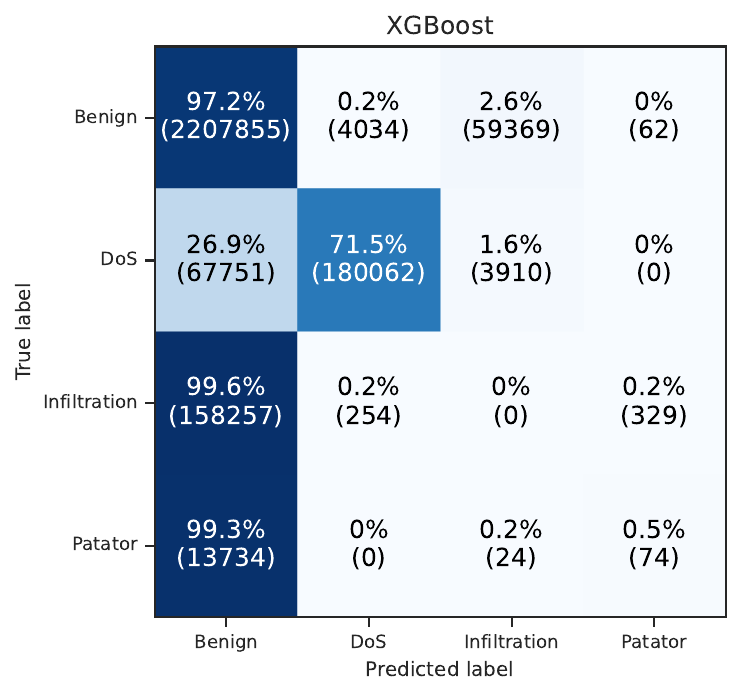}
		\caption{CIC-IDS2017 Multiclass}
		\label{fig:2017_cm_multi}
	\end{minipage}\hfill
	\begin{minipage}[t]{0.49\linewidth}
		\centering
		\includegraphics[width=\linewidth]{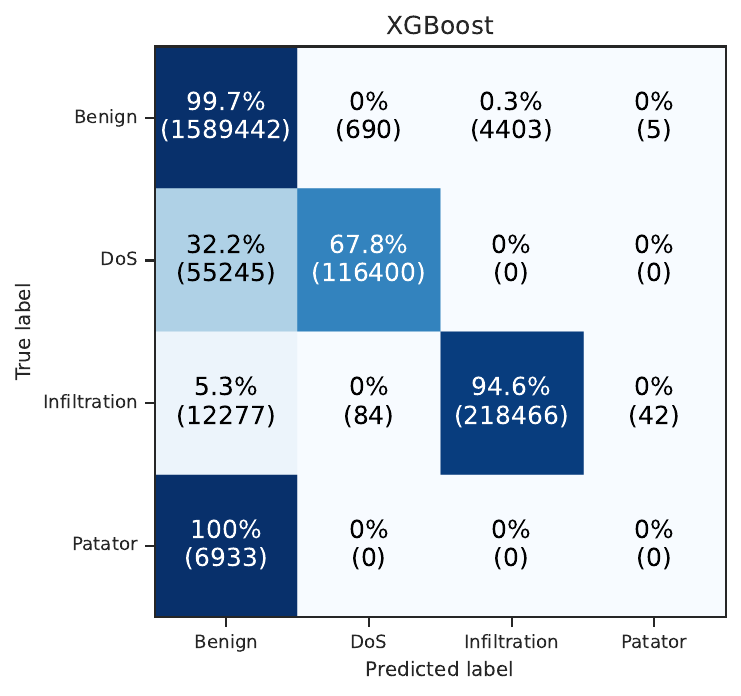}
		\caption{CNS-CIC-IDS2017 Multiclass}
		\label{fig:cns2017_cm_multi}
	\end{minipage}
\end{figure}

In contrast to UNSW-NB15, the CIC-IDS2017 dataset performs consistently worse in both binary and multiclass classification, as can be seen in \Cref{tab:results_eloocv_cic}. While this may seem discouraging, the drop in performance is itself revealing: it reinforces the idea that high scores obtained using naive data splits (e.g., 70/30 or 10-fold CV) are likely overestimations caused by information leakage.
In binary classification, even the best-performing models drop from near-perfect f1-scores of 0.997 (under 70/30 splits) to a much more grounded 0.596 when evaluated using \gls{eloocv}. Multiclass classification suffers even more severely, with f1-scores falling below 0.51 and balanced accuracy values going as low as 24\% for \gls{rf}.
These results highlight an important point: under more realistic deployment conditions, these models would fail to detect most attacks. For instance, the confusion matrix in \Cref{fig:2017_cm_multi} shows that the multiclass model fails to detect even a single sample of the \textit{Infiltration} class and classifies only 0.5\% of \textit{Patator} samples correctly.

Interestingly, the \enquote{corrected} version of the dataset (that fixed some underlying issues in the data), CNS-IDS2017, shows a notable improvement across all metrics, as can be seen in \Cref{tab:results_eloocv_cic}. This supports the notion that flaws in the original dataset construction (e.g., poor labeling or flow parsing issues) significantly impacted model generalizability. Binary classification reaches an F1-score of 0.949 and balanced accuracy of 95\% with \gls{xgboost}. Multiclass performance also improves, with the model correctly classifying most \textit{Infiltration} samples, while still struggling with classes such as \textit{Web-XSS} and \textit{Web-SQLi}, as shown in \Cref{fig:cns2017_cm_multi}.
\begin{table}[!t]
	\centering
	\caption{\gls{eloocv} results by dataset and maximum tree depth. BA is balanced accuracy, P is precision, and R is recall.}
	\label{tab:results_eloocv_cic}
	\footnotesize
	\setlength{\tabcolsep}{1.65pt}
	\begin{tabular}{@{}lr rrrr @{\hspace{5pt}} rrrr @{\hspace{5pt}} rrrr @{}}
		\toprule
		& & \multicolumn{4}{c}{CIC-IDS2017} & \multicolumn{4}{c}{CNS-IDS2017} & \multicolumn{4}{c}{CIC-IDS2018} \\
		\cmidrule(lr){3-6}\cmidrule(lr){7-10}\cmidrule(lr){11-14}
		Model & Depth & BA & P & R & F1 & BA & P & R & F1 & BA & P & R & F1 \\
		\midrule
		\multicolumn{14}{@{}l}{\textit{Binary classification}} \\
		\gls{rf} & 3 & 0.554 & 0.275 & 0.303 & 0.289 & 0.932 & 0.922 & 0.888 & 0.905 & 0.772 & 0.870 & 0.686 & 0.767 \\
		 & 5 & 0.643 & 0.573 & 0.351 & 0.435 & 0.881 & 0.962 & 0.771 & 0.856 & \textbf{0.790} & \textbf{0.930} & \textbf{0.647} & \textbf{0.763} \\
		 & 7 & 0.647 & 0.663 & 0.337 & 0.447 & 0.882 & 0.976 & 0.769 & 0.861 & 0.754 & 0.972 & 0.530 & 0.686 \\
		 & 10 & 0.680 & 0.813 & 0.382 & 0.519 & 0.832 & 0.988 & 0.667 & 0.797 & 0.746 & 0.984 & 0.502 & 0.665 \\
		\addlinespace[2pt]
		\gls{xgboost} & 3 & \textbf{0.723} & \textbf{0.801} & \textbf{0.474} & \textbf{0.596} & \textbf{0.954} & \textbf{0.989} & \textbf{0.912} & \textbf{0.949} & 0.680 & 0.896 & 0.429 & 0.580 \\
		 & 5 & 0.719 & 0.791 & 0.469 & 0.589 & 0.903 & 0.988 & 0.809 & 0.889 & 0.653 & 0.930 & 0.341 & 0.499 \\
		 & 7 & 0.717 & 0.797 & 0.463 & 0.586 & 0.894 & 0.988 & 0.792 & 0.879 & 0.645 & 0.958 & 0.308 & 0.466 \\
		 & 10 & 0.713 & 0.778 & 0.458 & 0.577 & 0.882 & 0.987 & 0.767 & 0.863 & 0.654 & 0.963 & 0.325 & 0.485 \\
		\midrule
		\multicolumn{14}{@{}l}{\textit{Multiclass classification}} \\
		\gls{rf} & 3 & 0.241 & 0.045 & 0.034 & 0.038 & 0.608 & 0.982 & 0.758 & 0.826 & 0.179 & 0.256 & 0.219 & 0.236 \\
		 & 5 & 0.261 & 0.134 & 0.044 & 0.067 & 0.568 & 0.984 & 0.690 & 0.754 & 0.163 & 0.429 & 0.068 & 0.101 \\
		 & 7 & 0.259 & 0.209 & 0.027 & 0.048 & 0.477 & 0.973 & 0.513 & 0.570 & 0.161 & 0.342 & 0.022 & 0.040 \\
		 & 10 & 0.267 & 0.389 & 0.046 & 0.082 & 0.470 & 0.920 & 0.497 & 0.532 & 0.300 & 0.821 & 0.065 & 0.114 \\
		\addlinespace[2pt]
		\gls{xgboost} & 3 & \textbf{0.423} & \textbf{0.599} & \textbf{0.439} & \textbf{0.506} & \textbf{0.655} & \textbf{0.986} & \textbf{0.832} & \textbf{0.896} & 0.544 & 0.696 & 0.250 & 0.288 \\
		 & 5 & 0.402 & 0.601 & 0.349 & 0.441 & 0.595 & 0.982 & 0.701 & 0.816 & \textbf{0.573} & \textbf{0.705} & \textbf{0.253} & \textbf{0.284} \\
		 & 7 & 0.391 & 0.600 & 0.256 & 0.359 & 0.617 & 0.982 & 0.752 & 0.847 & 0.572 & 0.803 & 0.255 & 0.286 \\
		 & 10 & 0.375 & 0.602 & 0.256 & 0.359 & 0.599 & 0.982 & 0.716 & 0.823 & 0.571 & 0.802 & 0.254 & 0.284 \\
		\bottomrule
	\end{tabular}
\end{table}

\subsubsection*{CIC-IDS2018}
On CIC-IDS2018, binary classifiers perform somewhat better than on CIC-IDS2017, particularly for \gls{rf}, which achieves balanced accuracy values over 75\%, as can be seen in \Cref{tab:results_eloocv_cic}. \gls{xgboost} lags behind, especially in recall, limiting its usefulness for detection in practical scenarios. Despite the improvements, even the best binary models show recall scores that remain insufficient for effective intrusion detection, with some models missing over 30\% of attacks.

The performance degrades further in the multiclass setting, where both balanced accuracy and recall drop substantially across the board. Even the best-performing models are only able to identify around 25\% of attack samples, and among those labeled as attacks, only 70–80\% are actually true attacks. These are critical shortcomings, as a high \gls{fnr} limits the usefulness of any intrusion detection system.

An especially unexpected result is the relatively high recall for \gls{xss} attacks, despite the fact that CIC-IDS2018 is a flow-based dataset and thus lacks payload-level features. Since \gls{xss} attacks are typically identified by patterns in HTTP body content (e.g., suspicious strings in form inputs), this raises questions about what characteristics the model is actually learning. We hypothesize that these results are likely due to unintended biases or artifacts introduced during the dataset generation process such as, differences in packet timing, flow lengths, or byte distributions that may have been inadvertently influenced by how \gls{xss} attacks were simulated. If that is the case, then the classifier is learning to identify an artifact of the environment or tooling used to generate the data, not the attack itself. This means the model would likely fail to detect real-world \gls{xss} attacks, especially when executed with different tools or different environments. This highlights a fundamental limitation of applying ML to these (and similar) tasks: models may learn correlations that yield high performance under a given validation setup but do not reflect realistic or generalizable behavior. As a result, particular care is required both in the design of synthetic datasets and in the interpretation of results derived from them.

\begin{figure}[!htbp]
	\centering
	\includegraphics[width=0.70\linewidth]{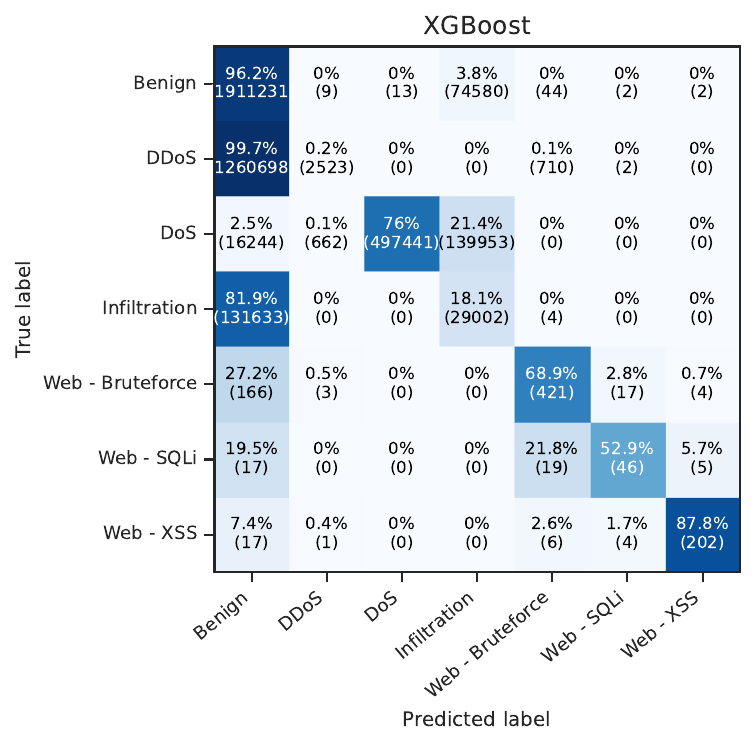}
	\caption{CIC-IDS2018 Multiclass}
	\label{fig:2018_cm_multi}
\end{figure}

\subsubsection*{Expert models}
We hypothesized that isolating each attack type by re-labeling all other samples as non-attack would simplify the classification task by reducing the complexity of the decision boundary. Each resulting model targets one attack category and treats all unrelated traffic as negative.

Our results strongly support this hypothesis. In UNSW-NB15, all but one attack class reached recall scores above 95\%, indicating that expert models are highly effective at recognizing their target class.

Compared with the CIC-IDS2018 multiclass results in \Cref{fig:2018_cm_multi}, the \gls{xgboost} expert raised \textit{DDoS} recall from 0.2\% to 68.4\%, with 50.3\% precision. The \gls{rf} expert increased \textit{Infiltration} recall from 18.1\% to 58.7\%.

Running multiple experts in parallel adds deployment complexity. The \textit{DDoS} result shows why both missed detections and false alarms matter when assessing this trade-off.

In operational settings, this structure offers further advantages. By isolating the detection task, each model can be paired with tailored countermeasures specific to the attack type. For instance, flows flagged as \textit{Patator} could trigger credential lockouts, while \textit{DDoS} detections might prompt temporary rate-limiting or connection drops.

Operators could route conflicting alerts for manual inspection. We evaluated individual experts, so whether combining alerts improves confidence or reduces false alarms remains open.

\subsection{Evaluating Generalization}\label{sec:evaluating_generalization}
Several works (e.g., \cite{Catillo2023}) highlight concerns about the poor transferability of \gls{ml}-based intrusion detection models. To assess this, we evaluated cross-dataset generalization using CIC-IDS2017 and CIC-IDS2018, which share features and attack types but differ in topology and environment. Models were trained on one dataset and tested exclusively on the other to avoid sample or temporal overlap.

As shown in \Cref{tab:results_multidataset}, performance drops sharply in cross-dataset settings. Binary classifiers rarely achieve recall above 70\%, with substantial variability across algorithms. Precision is similarly unstable: it remains relatively high when training on CIC-IDS2017 but degrades markedly when models are trained on CIC-IDS2018. This highlights a key challenge for ML-based IDS: even when attacks are generated using largely the same tools, models exhibit limited transferability. Although the experimental environment evolved from CIC-IDS2017 to CIC-IDS2018, these results support earlier findings \cite{Catillo2023} that models trained in one setting struggle to adapt to even modest changes in operating conditions—an increasingly common scenario in cloud and hybrid infrastructures. In practice, low recall corresponds to missed attacks, while low precision results in excessive false alarms that can overwhelm operators.

Frequent retraining may help but depends on labeled data, adds operational overhead, and offers limited benefit if datasets remain narrow or biased. As a result, current \gls{ml}-based \gls{ids} lack robustness for long-term or cross-domain deployment. Class imbalance, inconsistent labels, and synthetic traffic further limit robustness.

\begin{table}[!t]
	\centering
	\small
	\setlength{\tabcolsep}{2pt} 
		\caption{Experimental results using different datasets for training and testing}
	\begin{tabular}{@{}l l l >{\raggedleft}p{4.3em} rrrr@{}}
		\rotatebox{90}{Training} & \rotatebox{90}{Testing} & Binary & Algorithm & Balanced Acc. & Precision & Recall & F1-Score \\ 
		\midrule
		\multirow{6}{*}{\rotatebox{90}{2017} } & \multirow{6}{*}{\rotatebox{90}{2018} } 
		& Yes & AdaBoost & 0.648 & 0.877 & 0.367 & 0.517 \\
		& & & \textbf{\gls{rf}}  & \textbf{0.787} & \textbf{0.897} & \textbf{0.683} & \textbf{0.776} \\
		& & & \gls{xgboost}   & 0.617 & 0.996 & 0.235 & 0.381 \\ \cmidrule{3-8}
		& & No  & AdaBoost & 0.136 & 0.341 & 0.207 & 0.180 \\
		& & & \gls{rf}  & 0.163 & 0.309 & 0.486 & 0.376 \\
		& & & \textbf{\gls{xgboost}}   & \textbf{0.174} & \textbf{0.408} & \textbf{0.513} & \textbf{0.392} \\ \cmidrule{1-8}
		\multirow{6}{*}{\rotatebox{90}{2018} } & \multirow{6}{*}{\rotatebox{90}{2017} }
		& Yes & AdaBoost & 0.656 & 0.325 & 0.633 & 0.430 \\
		& & & \textbf{\gls{rf}}  & \textbf{0.658} & \textbf{0.538} & \textbf{0.401} & \textbf{0.459} \\
		& & & \gls{xgboost}   & 0.543 & 0.256 & 0.295 & 0.274 \\ \cmidrule{3-8}
		& & No  & AdaBoost & 0.147 & 0.623 & 0.426 & 0.479 \\
		& & & \gls{rf}  & 0.197 & 0.732 & 0.800 & 0.759 \\
		& & & \textbf{\gls{xgboost}}   & \textbf{0.286} & \textbf{0.800} & \textbf{0.841} & \textbf{0.813} \\
	\end{tabular}

	\label{tab:results_multidataset}
\end{table}
\section{Discussion}
The literature on \gls{ml}-based intrusion detection often reports near-perfect metrics, but such claims are undermined by methodological flaws: missing implementation details, disregard for class imbalance, and data leakage caused by improper partitioning. By applying the stricter \gls{eloocv} protocol, which enforces clean separation between training and testing flows, we observed significant performance drops, especially in multiclass tasks. Binary models retained higher scores, but their operational value is limited since they only flag anomalous traffic without providing attack context.  

Multiclass classifiers, although more informative, proved unreliable: rare attacks were often misclassified or ignored altogether. This exposes how misleading previously reported metrics can be under naive evaluation. Dataset quality also plays a decisive role: corrected versions of CIC-IDS~\cite{Liu2022} produced markedly better results than the originals~\cite{Sharafaldin2018}, indicating that many optimistic claims in prior work stem from flawed or noisy inputs.  

Cross-dataset validation further highlighted weak generalization. Even with similar features and attack types, models trained on one dataset performed poorly on another, echoing the concerns of~\cite{Catillo2023}. This lack of robustness limits deployment in continuously changing networks, particularly in cloud and hybrid environments.  


\subsubsection{Lessons Learned} This study highlights several practical lessons for evaluating ML-based flow intrusion detection systems using public datasets. First, \textbf{shuffled or randomly stratified splits should be avoided}, as they can introduce substantial information leakage due to correlated flows generated by the same attack activity. \textbf{Group-based or time-aware partitioning strategies} are more appropriate for this domain and should be clearly described to ensure reproducibility. When such information is not available, the dataset itself may be ill-suited for rigorous evaluation; experimental validity should not be compromised to accommodate dataset limitations. Second, \textbf{aggregate performance metrics alone are insufficient} to assess detection effectiveness in highly imbalanced settings. Per-class recall, macro-averaged metrics, and explicit reporting of class distributions or campaign counts are necessary to expose failures on minority attacks that are otherwise hidden by optimistic averages.

Evaluation on a single dataset provides limited insight into robustness. Even closely related benchmarks can differ substantially, and \textbf{cross-dataset or cross-scenario validation is essential to assess generalization}. Finally, \textbf{expert models can improve detection} for specific attack types, but they also \textbf{introduce additional deployment complexity}. Their use should therefore be framed as a diagnostic or complementary strategy, accompanied by discussion of scalability and integration within broader detection architectures rather than as standalone solutions.

\section{Threats to Validity}
\label{chap:threats}
\subsubsection{Internal Validity}
This is primarily affected by how attack instances are separated during evaluation. Because public intrusion detection datasets do not provide explicit campaign identifiers, attack separation is approximated using day-based grouping. To preserve closed-set supervised learning under strict partitioning, limited label coalescing was applied when attack categories occurred on a single day and would otherwise be absent from the training set. This intervention affects only a restricted subset of attack categories, namely the DoS and Patator families in CIC-IDS2017 and the DoS and DDoS classes in CIC-IDS2018 (as was also done in other related works, even if without the same motivation); all other attack types, including those in UNSW-NB15, infiltration attacks, and web-based attacks, are unaffected. In these cases, models are required to generalize to attack instances generated by tools not observed during training, effectively treating tool-specific variants as unseen instances within the same attack category. In any case, CICFlowMeter extracts only flow-level statistical features without application-layer semantics, so tool-specific differences are not preserved. 
Even if multi-day executions were available, separating these tools into distinct labels would result in highly imbalanced and weakly separable subclasses driven primarily by tool-specific artifacts rather than meaningful behavioral distinctions. Overall, these choices introduce conservative evaluation conditions and do not compromise the validity of the experimental conclusions.

\subsubsection{Construct Validity}

Construct validity is constrained by the use of flow-level features, which abstract away payload and protocol semantics. As a result, detection performance may reflect dataset artifacts or traffic generation biases rather than intrinsic attack behavior, particularly for application-layer attacks such as XSS or SQL injection. Also, no single evaluation metric fully captures operational IDS effectiveness under severe class imbalance. While multiple metrics are reported, our emphasis on recall and balanced accuracy reflects a security-centric perspective. Binary, multiclass, and expert task formulations provide complementary but incomplete views of detection capability; none alone represents real deployments, but together they expose recurring failure modes that are often hidden by aggregate metrics.

\subsubsection{External Validity}

Reliance on public benchmark datasets, which contain synthetic traffic, tool-specific artifacts, and constrained environmental diversity limits external validity. While this enables comparison with prior work, results may not transfer to operational networks with different workloads, topologies, or attacker behavior. Cross-dataset generalization experiments are also restricted to overlapping features and labels, simplifying real-world migration scenarios where both evolve over time. Moreover, static datasets cannot capture concept drift or adaptive adversaries, meaning that real-world performance is likely to degrade further without continuous retraining and explicit drift-handling mechanisms.

\section{Conclusion}
\label{chap:conclusion}

This practical experience report shows that many near-perfect results reported for ML-based intrusion detection are largely artifacts of flawed evaluation practices, most notably information leakage in flow-based datasets. When evaluated under leakage-resistant, event-level partitioning, model performance drops substantially: generalization is limited, multiclass detection degrades sharply, and robustness across datasets remains weak. These findings indicate that the practical detection capability of current ML-based IDS, at least when assessed using widely adopted public flow datasets, is far more constrained than suggested by optimistic benchmark results.

Rather than advocating increasingly complex models, the lessons from this work highlight the importance of careful evaluation design, transparent reporting, and realistic interpretation of results. Progress in ML-based intrusion detection depends critically on validation strategies that respect the structure of attack data, on metrics that expose failures on minority attacks, and on explicit assessment of generalization beyond a single dataset.

Future work should prioritize better-curated, well-documented datasets and standardized, leakage-resistant validation across datasets, scenarios, and time. It should also test whether packet-level, temporal, and unsupervised approaches address the limitations of flow-based evaluation.

\begin{credits}
\subsubsection{\ackname} The author thanks Rodrigo Machado for his MSc research, which supported the experimental work reported in this paper.
\end{credits}

\clearpage
\bibliographystyle{splncs04}
\bibliography{references}

\end{document}

%% file: glossary.tex
\newacronym{ae}{AE}{Autoencoder}
\newacronym{ai}{AI}{Artificial Intelligence}
\newacronym{aids}{AIDS}{Anomaly-based Intrusion Detection System}
\newacronym{cia}{CIA}{Confidentiality, Integrity, and Availability}
\newacronym{cm}{CM}{Confusion Matrix}
\newacronym{cnn}{CNN}{Convolutional Neural Network}
\newacronym{cv}{CV}{Cross-Validation}
\newacronym{dbscan}{DBSCAN}{Density-Based Spatial Clustering of Applications with Noise}
\newacronym{ddos}{DDoS}{Distributed Denial of Service}
\newacronym{dl}{DL}{Deep Learning}
\newacronym{dnn}{DNN}{Deep Neural Network}
\newacronym{dos}{DoS}{Denial of Service}
\newacronym{dt}{DT}{Decision Tree}
\newacronym{eda}{EDA}{Exploratory Data Analysis}
\newacronym{eloocv}{ELOOCV}{Experiment-wise Leave-One-Out Cross-Validation}
\newacronym{far}{FAR}{False Alarm Rate}
\newacronym{fips}{FIPS}{Federal Information Processing Standards}
\newacronym{fnr}{FNR}{False Negative Rate}
\newacronym{fpr}{FPR}{False Positive Rate}
\newacronym{hc}{HC}{Hierarchical Clustering}
\newacronym{hids}{HIDS}{Host-based Intrusion Detection System}
\newacronym{idps}{IDPS}{Intrusion Detection and Prevention System}
\newacronym{ids}{IDS}{Intrusion Detection System}
\newacronym{ips}{IPS}{Intrusion Protection System}
\newacronym{lr}{LR}{Logistic Regression}
\newacronym{lstm}{LSTM}{Long Short-Term Memory}
\newacronym{loocv}{LOOCV}{Leave-One-Out Cross-Validation}
\newacronym{ml}{ML}{Machine Learning}
\newacronym{mlp}{MLP}{Multilayer Perceptron}
\newacronym{mse}{MSE}{Mean Squared Error}
\newacronym{nids}{NIDS}{Network-based Intrusion Detection System}
\newacronym{nlp}{NLP}{Natural Language Processing}
\newacronym{nn}{NN}{Neural Network}
\newacronym{owasp}{OWASP}{Open Worldwide Application Security Project}
\newacronym{pca}{PCA}{Principal Component Analysis}
\newacronym{pii}{PII}{Personally Identifiable Information}
\newacronym{relu}{ReLU}{Rectified Linear Unit}
\newacronym{rf}{RF}{Random Forest}
\newacronym{roc}{ROC}{Receiver Operator Characteristics}
\newacronym{sdlc}{SDLC}{Software Development Lifecycle}
\newacronym{sids}{SIDS}{Signature-based Intrusion Detection System}
\newacronym{siem}{SIEM}{Security Information and Event Management}
\newacronym{sqli}{SQLi}{SQL injection}
\newacronym{svm}{SVM}{Support Vector Machine}
\newacronym{tnr}{TNR}{True Negative Rate}
\newacronym{tpr}{TPR}{True Positive Rate}
\newacronym{vae}{VAE}{Variational Autoencoder}
\newacronym{waf}{WAF}{Web Application Firewall}
\newacronym{xgboost}{XGBoost}{eXtreme Gradient Boosting}
\newacronym{xss}{XSS}{Cross-site Scripting}

\newacronym{ELOOCV}{ELOOCV}{Experiment-wise Leave-one-out Cross-Validation}

\newacronym{pr}{PR}{Precision-Recall}

\newacronym{aucpr}{AUCPR}{Area Under the Precision-Recall Curve}